# Vertical spin transport in Al with Pd/Al/Ni$_{80}$Fe$_{20}$ trilayer films at room temperature by spin pumping

Yuta Kitamura, Eiji Shikoh, Yuichiro Ando, Teruya Shinjo, Masashi Shiraishi*

*Graduate School of Engineering Science, Osaka University, 1-3 Machikaneyama-cho, Toyonaka, Osaka 560-8531, Japan.*

*Corresponding author: shiraishi@ee.es.osaka-u.ac.jp

**Spin pumping enables the vertical transport of pure spin current through Al in a Pd/Al/Ni$_{80}$Fe$_{20}$(Py) trilayer film, in which the Py acts as a spin battery. The spin current injected into the Al flows through the Al to reach the Pd, resulting in the generation of electromotive forces due to the inverse spin Hall effect in the Pd. The electromotive forces decreased with increasing thickness of the Al layer. A simple model based on the theory by Tserkovnyak et al., allows an estimation of the spin coherence of the vertical spin transport in the Al of 61 nm. This comparatively short coherence is attributed to a reduction in spin pumping efficiency because of the roughness of the Al/Py interface.**

Since spin-polarized electrons flowing in solids show intriguing characteristics in association with spin transfer, spin transport, and spin accumulation,[1] the physical properties of flowing spins are actively studied as 'spintronics'. In the field of spintronics, spin pumping[2-4] has attracted considerable attention as a method of dynamical spin injection, which induces a pure spin current in a nonmagnetic metal.[5] Concurrently, the inverse-spin Hall effect (ISHE) in paramagnets, which enables the conversion from a pure spin current to a charge current, is frequently used to detect a pure spin current.[5,6] Although the number of studies of the transport of a pure spin current (not a spin-wave spin current) in a spacer between a ferromagnet and a paramagnet using the dynamical method remains limited,[7,8] previous studies have shown that spin pumping has the potential to generate a normal pure spin current in a wide range of materials. The previous demonstrations showed spin transport in lateral structures by dynamical spin injection.[7,8] However, it is significant that dynamical spin transport is studied in a device with various structures that enables spin transport in various materials, such as molecules,.

In this work, we demonstrate successful perpendicular spin transport in the Al layer of a Pd/Al/Ni$_{80}$Fe$_{20}$(Py) trilayer film by the spin pumping at room temperature (RT). The spin transport characteristics are investigated using a simple model based on the theory by Tserkovnyak et al.[4] The estimated spin coherence length in the Al was comparatively shorter (61 nm) than that observed using a lateral spin valve structure (ca. 300 nm at RT[9,10]). In this paper, we clarify the effect of inhomogeneity of the Al/Py interface on the spin injection efficiency, which results in a shorter spin coherence length.

**Result**

**Sample structure and inhomogeneity of the Al/Py interface.** A schematic structure of the sample used in this study is shown in Fig. 1(a). The sample consisted of a Pd/Al/Py trilayer film with lateral dimensions of 1 mm×1.9 mm ($l \times w$). The thicknesses of the Pd ($d_{N2}$) and Py ($d_{FM}$) layers were 15 nm, and that of the Al ($d_{N1}$) was varied from 5 to 40 nm. The electrical conductivity of the Py, Al, and Pd were $\sigma_{FM}$=2.05×10$^6$ (Ωm)$^{-1}$, $\sigma_{N1}$=9.42×10$^6$ (Ωm)$^{-1}$, and $\sigma_{N2}$=4.08×10$^6$ (Ωm)$^{-1}$, respectively, according to four-terminal measurements. Note that we fabricated samples by two different methods: (A) The Pd, Al, and Py layers were formed with a substrate cooling system and a deposition rate of 0.25 nm/sec, and (B) the Pd, Al, and Py layers were formed without substrate cooling at a deposition rate of 0.04 nm/sec. In the following discussion, these two sample types are referred to as sample (A) and sample (B). By using the substrate cooling system, the formation of clusters could be suppressed. Additionally, in the case of the vacuum thermal evaporation, the faster the deposition rate, the more effectively the formation of the clusters is suppressed, as we demonstrate in the following section. Therefore, by comparing the results for samples (A) and (B) we can discuss the effect of the Al/Py interfacial roughness, which can have a non-negligible effect on the spin injection efficiency.

**Dynamical spin injection and inverse spin Hall effect.** Whereas a pure spin current can influence the magnetization of a ferromagnet,[11,12] precessing magnetization loses torque by emitting a spin current.[2] In a ferromagnetic resonance (FMR) state, the magnetization precession around a static external magnetic field enables spin pumping, transferring spin angular momentum from the ferromagnet to an adjacent nonmagnetic layer. Therefore, spin pumping is a reciprocal process of spin-current-induced magnetization switching.[11,12] The dynamics of the moving magnetization $\mathbf{M}(t)=\mathbf{M}+\mathbf{m}(t)$ in the ferromagnet layer are described by the Landau-Lifshitz-Gilbert equation as,[13]

$$\frac{d\mathbf{M}(t)}{dt} = -\gamma \mathbf{M}(t) \times \mathbf{H}_{\text{eff}} + \frac{\alpha}{M_S}\mathbf{M}(t) \times \frac{d\mathbf{M}(t)}{dt}, \tag{1}$$

where **M**($t$), **M**, **m**($t$), $\gamma$, **H**$_{\text{eff}}$, $M_S$, and $\alpha$ are the magnetization, the static component of the magnetization, the dynamical component of the magnetization, the gyromagnetic ratio, the effective magnetic field, the saturation magnetization, and the dimensionless Gilbert damping constant, respectively. Because of the second term on the right side of equation (1), the precession spirals down to a static magnetization axis by the damping effect. In addition, the damping constant becomes large when the emission of spin angular momentum from the ferromagnet to a connected nonmagnet occurs, i.e., after dynamical spin injection into a nonmagnet.[4] The enhancement of the damping effect has been theoretically[4] and experimentally[14] studied, and serves as a good index for the estimation of the spin current density injected into a nonmagnet by spin pumping. In the case of Py, Py has a very small magnetocrystalline anisotropy, so the magnetization in the Py was aligned along the film plane when the external magnetic field **H** was applied in the film plane. When the **H** and the magnetization precession frequency satisfy the FMR condition,[15] the spins parallel to the magnetization precession axis are pumped into the Al layer.[2]

The ISHE was used to detect the pure spin current.[5,6] The ISHE is a process that converts a pure spin current into a charge current, resulting in an electric voltage (a schematic image is shown as dashed arrows in Fig. 1(a)), and the spin current density is estimated from the amplitude of the ISHE voltage when the spin Hall angle is known. The charge current **J**$_C$ generated by the ISHE is expressed as,[16]

$$\mathbf{J}_C = \theta_{ISHE}\left(\frac{2e}{\hbar}\right)\mathbf{J}_S \times \sigma, \qquad (2)$$

where $\mathbf{J}_S$, $\theta_{ISHE}$, $e$, $\hbar$, and $\sigma$ are the spin current density, the spin Hall angle, the elementary charge, the Dirac constant, and the spin polarization vector of the spin current, respectively. By measuring the electromotive forces in the Pd under the FMR condition of the Py film, we can estimate the spin current density in the Pd layer.

Before starting the experiments, we measured the spin Hall angle of the Pd by the spin pumping method with a Pd/Py bilayer film to check the quality of the Pd electrode. The amplitude of the ISHE voltage at the Pd and the enhancement of the Gilbert damping constant were estimated to be 6.0 µV, and $7.3 \times 10^7$ s$^{-1}$, respectively. The spin Hall angle of the Pd was estimated to be 0.017 in our study, which agrees with the previously reported value[6] and indicates an acceptable sample quality.

**Observation of inverse spin Hall effect.** Figures 2(a) and (b) show FMR spectra d$I(H)$/d$H$ and the measured voltages, respectively, for the Pd/Al(20 nm)/Py film of sample (A) when an external magnetic field **H** was applied at $\theta = 0°$, 90°, and 180°, as shown in the right side of the Fig. 2(a). A clear electromotive force signal from the Pd layer at $\theta = 0°$ and 180° was observed around the FMR field, $H_{FMR}$ (111 mT). The amplitude of the ISHE voltage $V_{ISHE}$ can be extracted using the following equation,[5]

$$V(H) = V_{ISHE} \frac{\Gamma^2}{(H-H_{FMR})^2 + \Gamma^2} + V_{AHE} \frac{-2\Gamma(H-H_{FMR})}{(H-H_{FMR})^2 + \Gamma^2}, \qquad (3)$$

where $V(H)$, $\Gamma$, $V_{ISHE}$, and $V_{AHE}$ correspond to the output voltage, the damping constant, the amplitude of the ISHE, and the amplitude of the anomalous Hall effect (AHE), respectively. $V_{ISHE}(\theta=0°)$ and $V_{ISHE}(\theta=180°)$ were estimated to be 604 nV and -547 nV, respectively, using equation (3). Here, the electromotive force due to the ISHE changed sign when the external magnetic field was reversed. Therefore, we defined $V'_{ISHE}=(V_{ISHE}(\theta)-V_{ISHE}(\theta+180°))/2$ as an effective voltage for eliminating the heating effect, for which the polarity of the voltage does not change the sign. It is notable that the amplitude of the ISHE voltage disappeared when the external magnetic field was applied perpendicular to the film plane, although FMR of the Py was successful ($H_{FMR}(\theta=90°)$ was measured to be 1210 mT). This confirms the successful detection of the ISHE voltage in the Pd film.

Additional evidence for successful spin injection in the Al by the dynamical method was obtained. Figure 3(a) shows the effective output voltage $V'_{ISHE}$ from sample (A) with 20 nm of Al as a function of microwave power. The amplitude of the ISHE voltage is proportional to the microwave power (see Fig. 3(b)), which is reasonably explained by a direct-current-spin-pumping model and also indicates that the applied microwave power (200 mW) was lower than the saturation of the FMR absorption for this system.[17] The linearity of the amplitude of the ISHE voltage as a function of the microwave power and the inversion of the ISHE voltage polarity with magnetization reversal indicate successful dynamical spin injection and spin transport in the Al.

**Al-thickness dependence of physical parameters.** We quantitatively analyze the transport characteristics of the pure spin current in the Al. Figure 4(a) shows the Al-thickness dependence of the $V'_{ISHE}$ value estimated for the Pd/Al/Py films at a microwave excitation power of 200 mW. The solid red circles and blue crosses show $V'_{ISHE}$ of samples (A) and (B), respectively. For sample (A), $V'_{ISHE}$ decreases with increasing $d_{N1}$, because spin relaxation in the Al decreases the number of spins reaching the Pd film. Moreover, the decrease in the total resistance of the sample and the backflow of the spin current into the Py with increasing Al film thickness further decreased $V'_{ISHE}$. For sample (B), the amplitude of the ISHE voltages was small compared with those of sample (A) when the thickness of the Al was more than 20 nm.

The spin current injected into the Pd layer through the Al layer, $J^0_{S2}$, decays along the $y$ direction ($y=0$ is defined as a point on the Al/Pd interface. See Fig. 1(a).) due to spin relaxation, which we assume to be described by,[18]

$$J_{S2}(y) = \frac{\sinh[(d_{N2} - y)/\lambda_{N2}]}{\sinh(d_{N2}/\lambda_{N2})} J^0_{S2} , \qquad (4)$$

where $\lambda_{N2}$ is the spin diffusion length of the Pd. Since the charge current density $J_c(y)$ can be expressed as $J_C(y) = \theta_{ISHE}(2e/\hbar)J_{S2}(y)$, the average charge current density can be described as,[6]

$$\langle J_C \rangle = \theta_{ISHE}\left(\frac{2e}{\hbar}\right)\frac{\lambda_{N2}}{d_{N2}}\tanh\left(\frac{d_{N2}}{2\lambda_{N2}}\right)J^0_{S2}. \quad (5)$$

The total resistance of the trilayer can thus be obtained as,

$$V'_{ISHE} = \frac{wd_{N2}}{\sigma_F d_F + \sigma_{N1} d_{N1} + \sigma_{N2} d_{N2}}\langle J_C \rangle, \quad (6)$$

from an equivalent circuit of the Pd/Al/Py system, as shown in Fig. 1(b). Using equations (5) and (6), the spin current density injected into the Pd from the Al, $J^0_{S2}$, is given by,

$$J^0_{S2} = \frac{\sigma_F d_F + \sigma_{N1} d_{N1} + \sigma_{N2} d_{N2}}{w\theta_{ISHE}\lambda_{N2}\tanh(d_{N2}/2\lambda_{N2})}\left(\frac{\hbar}{2e}\right)V'_{ISHE}. \quad (7)$$

The spin current density $J^0_{S2}$ was estimated from $V'_{ISHE}$ by setting the parameter $\lambda_{N2}$ to 9 nm [19] from equation (7). Figure 4(b) shows the Al-thickness dependence of the spin current density, $J^0_{S2}$.

Meanwhile, the spin pumping generates a spin current from the precessing magnetization **M**(t), so the direct current component of a spin current in the Pd film can be expressed as,[6]

$$J^0_{S2} = \frac{\omega}{2\pi}\int_0^{2\pi/\omega}\frac{\hbar}{4\pi}g^{\uparrow\downarrow}_{eff}\frac{1}{M_S^2}\left[\mathbf{M}(t)\times\frac{d\mathbf{M}(t)}{dt}\right]_z dt, \quad (8)$$

where $\omega$ and $g^{\uparrow\downarrow}_{eff}$ are the angular frequency of the microwaves ($\omega=2\pi f$) and the real part of the effective mixing conductance in the Al layer in series with the two interfaces (one with Py and the other with Pd). The effective magnetic field $\mathbf{H}_{eff}$ can be described as $\mathbf{H}_{eff}(t)=\mathbf{H}+\mathbf{H}_m(t)+\mathbf{h}(t)$, where H=(0,0,H) is the external magnetic field, $\mathbf{H}_m(t)=[0,-4\pi m_y(t),0]$ is the dynamic demagnetization, and $\mathbf{h}(t)=(he^{i\omega t},0,0)$ is the external ac field. $m_y(t)$ and $h$ are the y component of **m**(t) and the amplitude of the external ac magnetic field, respectively. Under these conditions, the direct-current component of the spin current, $J^0_{S2}$, at the Pd/Al interface can be rewritten from equations (1) and (8) as,

$$J_{S2}^0 = \frac{g_{eff}^{\uparrow\downarrow} \gamma^2 h^2 \hbar \left[ 4\pi M_S \gamma + \sqrt{(4\pi M_S)^2 \gamma^2 + 4\omega^2} \right]}{8\pi \alpha^2 \left[ (4\pi M_S)^2 \gamma^2 + 4\omega^2 \right]}. \tag{9}$$

In this study, some of the pumped spins are scattered in the Al layer, some reach the Pd layer and are converted to a charge current, and the remainder diffuses back into the Py layer, which is different to the simple bilayer case. Here, the spins diffusing back into the Py layer return the spin angular momentum to the magnetization of the Py layer. Therefore, because of the backflow, the enhancement of the dimensionless Gilbert damping constant is reduced relative to the case of the Pd/Py bilayer,[4] which should be taken into account in our analysis. The enhancement of the dimensionless Gilbert damping constant due to spin relaxation in the trilayer system is given in MKS units as,[4]

$$\alpha(d_{N1}) = \frac{1}{\gamma M_S} \left( G_0 + \left[ 1 + g_{N1/FM}^{\uparrow\downarrow} \sqrt{\frac{3\tau_{SF}}{\tau_{el}}} \frac{\pi}{2k_F^2} \frac{1 + \tanh(d_{N1}/\lambda_{N1}) g_{N2/N1}^{\uparrow\downarrow} \sqrt{3\tau_{SF}/\tau_{el}} \left( \pi/2k_F^2 \right)}{\tanh(d_{N1}/\lambda_{N1}) + g_{N2/N1}^{\uparrow\downarrow} \sqrt{3\tau_{SF}/\tau_{el}} \left( \pi/2k_F^2 \right)} \right]^{-1}$$

$$\times \frac{(g_L \mu_B)^2}{4\pi \hbar} \frac{g_{N1/FM}^{\uparrow\downarrow}}{d_F} \times 10^7 \right), \tag{10}$$

where $G_0$, $g_{N1/FM}^{\uparrow\downarrow}$, $g_{N2/N1}^{\uparrow\downarrow}$, $\tau_{SF}$, $\tau_{EL}$, $k_F$, $\lambda_{N1}$, and $g_L$, are the bulk Gilbert damping constant of Py ($G_0 = \gamma M_S \alpha_0$), the real part of the mixing conductance of the Al/Py interface, the real part of the mixing conductance of the Pd/Al interface, the spin relaxation time, the elastic scattering time, the Fermi wavenumber, the spin diffusion length of the Al, and the g-factor of electrons, respectively. Setting $g_{N2/N1}^{\uparrow\downarrow} = 0$ decouples the Pd/Al junction, i.e., we can obtain the dimensionless Gilbert damping constant of the Al/Py bilayer. Here, the real part of the effective mixing conductance, $g_{eff}^{\uparrow\downarrow}$, for the Pd/Al/Py film can be determined from the enhancement of the dimensionless Gilbert damping constant due to the missing of the spin angular momentum during the spin pumping as,[20]

$$g_{eff}^{\uparrow\downarrow} = \frac{4\pi M_S d_F}{g_L \mu_B}\left(\alpha(d_{N1}) - \alpha(d_{N1})_{g_{N2/N1}^{\uparrow\downarrow}=0}\right). \tag{11}$$

Since the difference between the dimensionless Gilbert damping constants of Pd/Al/Py and Pd/Al depends on the presence of the Al layer, the effective mixing conductance also depends on the thickness of the Al layer. Using the amplitude of the observed ISHE voltages and equation (7), we obtain the spin current density injected into the Pd from the Al. Then, using equations (9), (10), and (11), the spin diffusion length of Al, $\lambda_{N1}$; the mixing conductances, $g_{N1/FM}^{\uparrow\downarrow}$ and $g_{N2/N1}^{\uparrow\downarrow}$; and the Gilbert damping constant of the bulk Py film, $G_0$ can be estimated.

Using equations (9)-(11) with $M_S$=0.065 T, $k_F$=1.75×10$^{10}$ m$^{-1}$, and $h$=6.0×10$^{-5}$ T and a spin flip probability $\varepsilon=\tau_{el}/\tau_{SF}$=1.5×10$^{-3}$,[21,22] we found $G_0$ = 7.8×10$^7$ s$^{-1}$, $g_{N1/FM}^{\uparrow\downarrow}$ = 4.0×10$^{20}$ m$^{-2}$, $g_{N2/N1}^{\uparrow\downarrow}$ = 4.0×10$^{19}$ m$^{-2}$, and $\lambda_{N1}$ = 61 nm for the Pd/Al/Py film (the fitting line is shown as a dashed line in Fig. 4(b)). Here, the sample data (Al thickness =15 nm) is not included for the fitting, since its spin current density was singularly large and the fitting calculation was not successfully completed, for reasons currently under investigation (we have confirmed that all of the experimental results are reproducible). The Gilbert damping constant of the bulk Py film, $G_0$, and the real part of the mixing conductance, $g_{N1/FM}^{\uparrow\downarrow}$, are close to the values reported in the literature,[23,24] and the bulk damping constant of the Py film was estimated to be $G_0 = 7 \times 10^7$ s$^{-1}$ on the basis of FMR measurements, and the real part of the mixing conductance of Al/Py was reported to be $g_{N1/FM}^{\uparrow\downarrow} = 2.0 \times 10^{20}$ m$^{-2}$. Although the real part of the mixing conductance of the Pd/Al has not yet been reported, the real part of the mixing conductance of Cu/Pt was reported to be $g_{N2/N1}^{\uparrow\downarrow} = 3.5 \times 10^{19}$ m$^{-2}$,[4] which is comparable to the value for our Pd/Al.

**Discussion**

The estimated spin diffusion length of the Al in this study is smaller than that estimated using an electrical method.[9,10,21,25] We assumed that this was due to the effect of interfacial roughness of the Al/Py interface on the spin injection efficiency, the following experiments were implemented. Figure 4(c) shows the Al-thickness dependence of the root mean square roughness R$_{RMS}$ values of the surface roughness of samples (A) and (B). The R$_{RMS}$ values of sample (A) are smaller than those of sample (B). This difference is especially obvious for thicker Al layers ($d_{N1}$>25 nm), which is ascribable to Al cluster formation induced by the sample temperature increase by heat radiation during the evaporation. It is notable that a dramatic decrease in spin current density was observed for samples with the same Al thickness ($d_{N1}$>25 nm, see Fig. 4(b)). This finding indicates that interfacial roughness between the Py and the Al impedes efficient spin injection. The effective magnetization, which was estimated from the FMR measurements, decreased with increasing Al thickness (see Fig.

4(d)), and this result strongly supports the above discussion and conclusions. Figures 4(e) and (f) show schematic illustrations of cross sectional views of a Py layers deposited on Al films of different roughness. For a sufficiently flat Al layer surface, such as that shown in Fig. 4(e), the Py film is uniformly magnetized along the in-plane direction. Meanwhile, for a rough Al layer surface, such as that shown in Fig. 4(f), the magnetization of the Py film is easily tilted from the in-plane direction,[26] which decreases the effective magnetization. The reduction of the effective magnetization is attributed to the roughness, and the rough interface impedes efficient spin injection through the interface because the spin current density is reduced due to the low effective magnetization. Hence, the spin injection efficiency decreases with increasing roughness of the Al surface. Our experimental results regarding the effective magnetization and roughness are in good agreement with the above interpretation. The findings in this study indicate that the control of the interface is of great significance in achieving effective dynamical spin injection. We expect that the combination of spin pumping and the ISHE can be used for spin injection, transport, and detection in various systems with vertical structures, such as molecular spin valves.

In conclusion, the Al thickness dependence of the amplitude of the ISHE voltage and the spin current density were studied using Pd/Al/Py trilayer films. The Gilbert damping constant of the bulk Py, the real part of the mixing conductance of the Al/Py, the real part of the mixing conductance of the Pd/Al, and the spin diffusion length in the Al were estimated to be $G_0 = 7.8 \times 10^7$ s$^{-1}$, $g_{N1/FM}^{\uparrow\downarrow}$ =

$4.0 \times 10^{20}$ m$^{-2}$, $g^{\uparrow\downarrow}_{N2/N1} = 4.0 \times 10^{19}$ m$^{-2}$, and $\lambda_{N1} = 61$ nm, respectively, by extending the conventional theoretical model. This estimation revealed the importance of interfacial control in efficient spin injection. The measurement and estimation methods provide a valuable technique for estimating the spin coherence of materials, where successful and reproducible spin injection and transport are awaited.

**Methods**

**Experimental procedure.** Two Au pads were formed as electrodes at both ends of the Pd (see Fig. 1(a)) by electron beam deposition using a shadow mask method. The Pd film, Al film, and Py film were formed on a thermally oxidized Si substrate by electron beam deposition using a shadow mask method with vacuum in-situ process. For sample (A), the temperature was kept at -2°C during the deposition by a cooling medium flowing through the sample holder. Two contacts for taking the inverse spin-Hall effect (ISHE) voltage measurements were attached to the Au Pads with Ag paste. The electromotive force due to the ISHE is obtained as $V_{ISHE}=[R_F R_{N1} R_{N2}/(R_F R_{N1}+R_{N1} R_{N2}+R_{N2} R_F)]I_C$ as shown in Fig. 1(b). $I_C=ld_{N2}\langle J_C \rangle$ is charge current generated by the ISHE. In the FMR measurements, each sample was placed at the center of a TE$_{102}$ microwave cavity in an electron spin resonance system (BRUKER, EMX10/12 EPR spectrometer) with a frequency of $f$=9.62 GHz at room temperature. In this system, the microwave amplitude $h$ was $6.0 \times 10^{-2}$ mT for a microwave power $P_{MW}$ of 200 mW. An external magnetic field **H** was applied at $\theta=0^\circ$, $90^\circ$, and $180^\circ$, as shown in Fig. 1(a). We used the resonance condition, $(\omega/\gamma)^2=H_{FMR}(H_{FMR}+4\pi M_S)$.

**Acknowledgments**

This study was partly supported by the Global COE program "Core Research and Engineering of Advanced Materials Interdisciplinary Education Center for Materials Research", by a Grant-in-Aid for Scientific Research from the MEXT, Japan, and by TORAY Science Foundation.

**Author contributions**

Y. K. fabricated the samples, collected all of the data, and performed analysis of the data. M. S, E. S., T.S. and Y. A. supported the experiments and the analysis of the data. Y. K., E. S. and M. S. wrote the manuscript. All authors discussed the results and commented on the manuscript.

**Additional information**

The authors declare no competing financial interests.

**Figure captions**

**Figure 1 | Sample structure and equivalent circuit model. (a)** A schematic illustration of the Pd(N1)/Al(N2)/Py(FM) trilayer film. **M** and **H** denote the precessing magnetization and the external magnetic field, respectively, and the magnetization dynamics of the Py induces spin pumping into the Al. The generated pure spin current flows in the Al, reaches the Pd, and is converted to a charge current in the Pd by the ISHE. The dashed arrows shown in the Pd layer describe spin scattering by the ISHE. **(b)** An equivalent circuit model for the Pd/Al/Py trilayer film.

**Figure 2 | Observation of the ISHE in sample (A). (a)** $H$ dependence of the FMR spectra, $dI(H)/dH$, measured for the Pd/Al(20 nm)/Py film under the microwave excitation of 200 mW. The angle of the external static magnetic field was set to be 0, 90, or 180 degrees, as shown at the right side of the graph. **(b)** $H$ dependence of the output voltages for the Pd/Al(20 nm)/Py film under the microwave excitation of 200 mW. The dashed lines show the theoretical fit obtained using equation (3).

**Figure 3 | Microwave power dependence of the electromotive force in the Pd. (a)** $H$ dependence of the effective output voltage $V'_{ISHE}$ for Pd/Al(20 nm)/Py film (sample (A)) under various microwave powers. **(b)** Microwave power dependence of $V'_{ISHE}$. The solid circles are the amplitude

of the ISHE voltages estimated using equation (3). The solid line shows a linear fit to the data.

**Figure 4 | Al-thickness dependence of physical parameters observed in this study.** The solid red circles and the blue crosses shown in (a)-(d) are the experimentally obtained values for samples (A) and (B), respectively. The error bars represent the standard error (three samples). **(a)** Al-thickness dependence of the amplitude of the ISHE voltages. **(b)** Al-thickness dependence of the spin current densities ($J^0_{S2}$), which reached the Pd through the Al. The dashed line shows the result of the theoretical fitting described in detail in the main text. **(c)** Al-thickness dependence of the $R_{RMS}$ values of the sample surface roughness. **(d)** Al-thickness dependence of the effective magnetization. The dotted line is a guideline showing $M_S$=0.065 T. **(e)** A schematic illustration of a cross sectional view of a Py layer deposited on a flat Al layer. **(f)** A schematic illustration of a cross sectional view of a Py layer deposited on a rough Al layer.

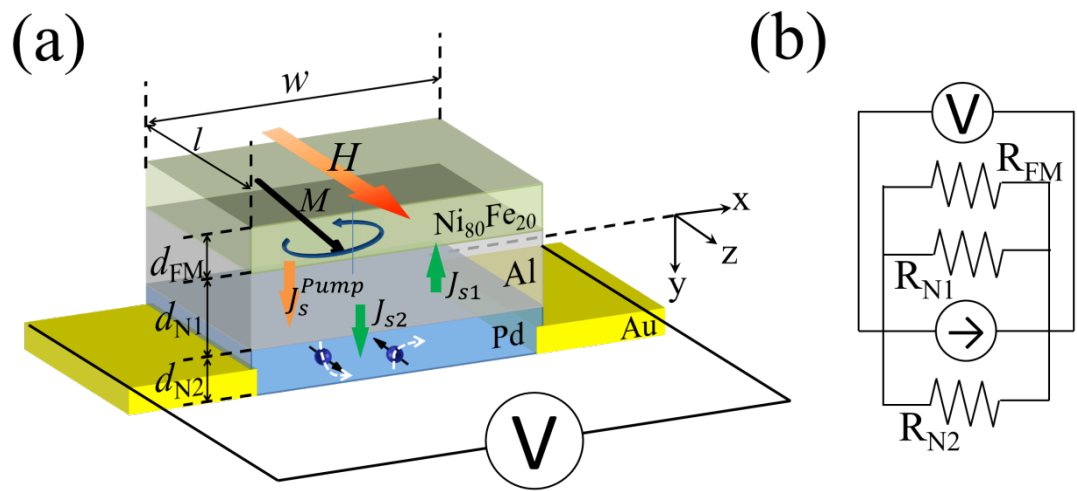

Kitamura, et. al., FIG. 1.

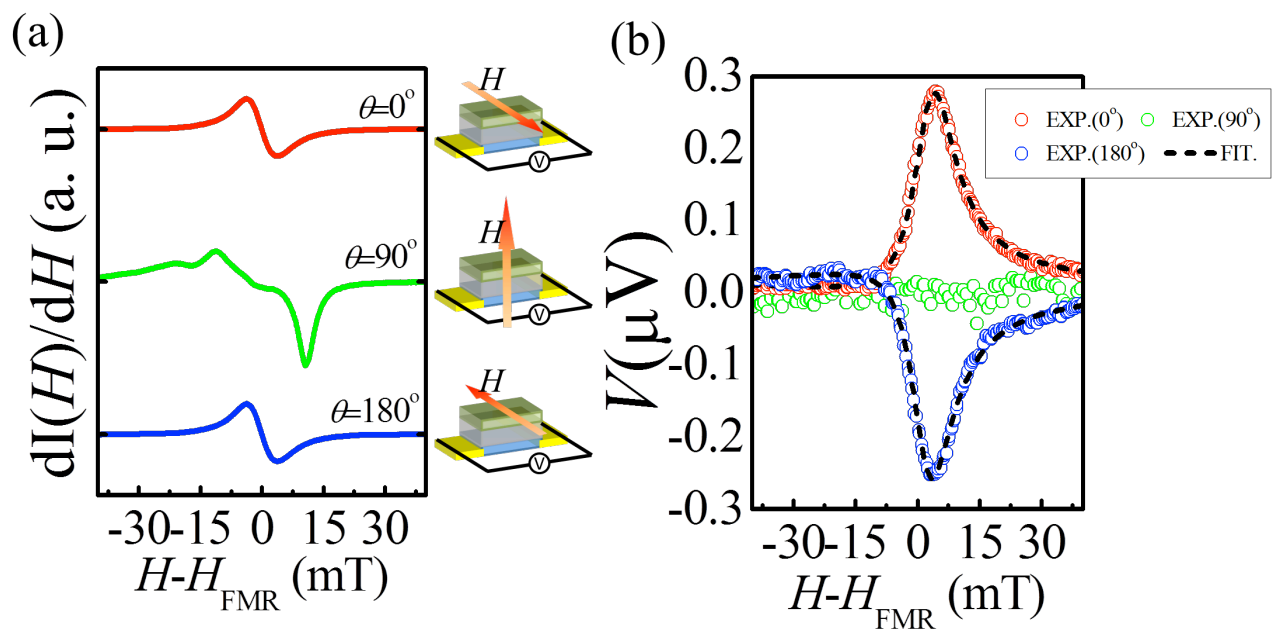

Kitamura, et. al., FIG. 2.

(a) 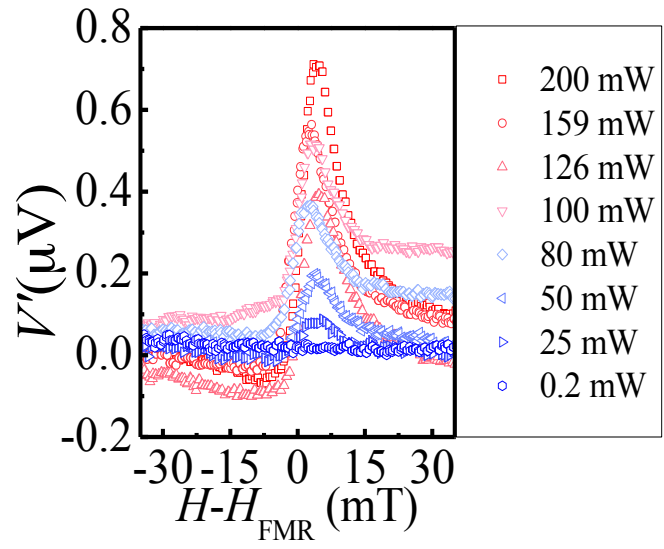 (b) 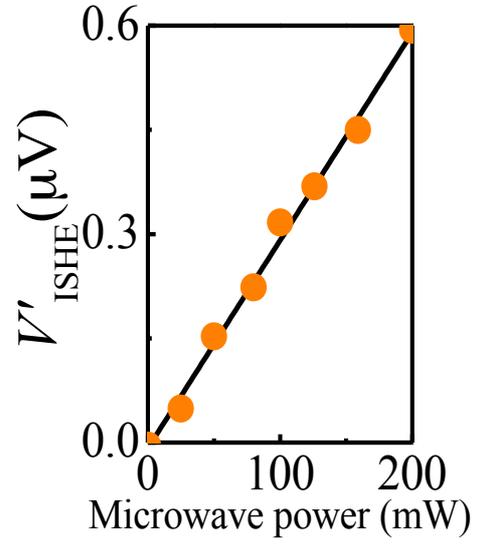

Kitamura, et. al., FIG. 3.

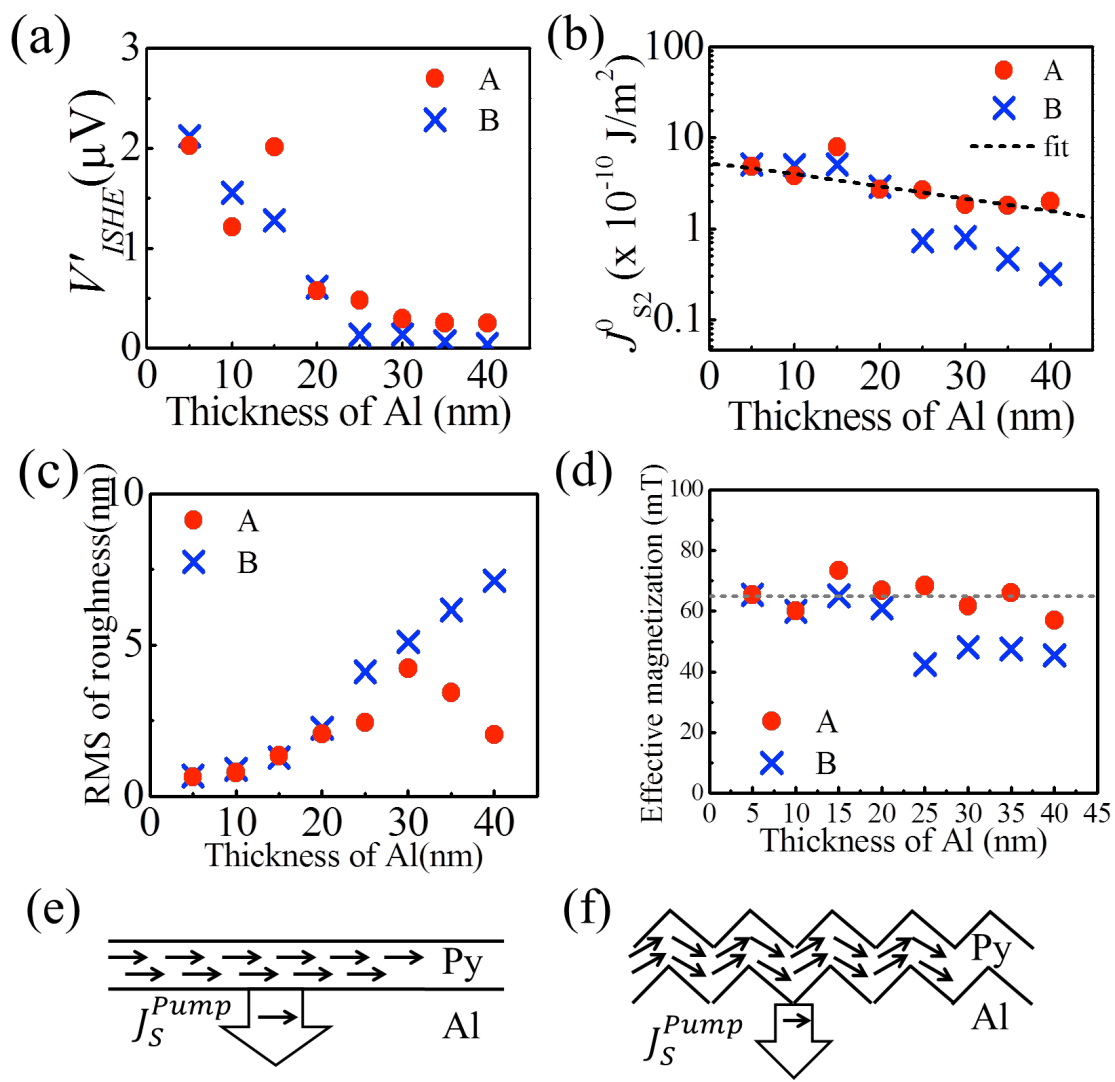

Kitamura, et. al., FIG. 4.